\documentclass[twocolumn]{article}
\usepackage[utf8]{inputenc}
\usepackage{booktabs}
\usepackage{dsfont}
\usepackage{graphicx}
\usepackage{hyperref}
\usepackage{mathrsfs}
\usepackage{listings}
\usepackage{xcolor}
\usepackage{enumerate}
\usepackage{amsmath,amssymb}

\title{Parallel Tempering for Logic Synthesis}

\author{Thomas H\"aner, Damian S.~Steiger, Helmut G.~Katzgraber}
\date{Amazon Quantum Solutions Lab}

\begin{document}

\maketitle

\abstract{The task of logic synthesis is to map a technology-independent representation of an application to hardware-specific operations, taking into account various constraints and trading off different costs associated with the implementation. Constraints may include the target gate library and feasible connectivity, whereas the costs capture, for example, the required chip area and delay. Here we propose to use parallel tempering Monte Carlo to find low-level implementations of a given Boolean function. In contrast to previous work leveraging Markov chain Monte Carlo methods such as simulated annealing, we do not start with an initially correct (but suboptimal) implementation that is then further optimized. Instead, our approach starts with a random logic network that is subsequently modified in order to reduce the error to zero. We apply our method to the task of synthesizing Majority-$n$ in terms of Majority-$3$ gates with and without inverters, aiming for a low Majority-$3$ count. Our method is able to find solutions that are on par or better than previously known solutions. For example, for $n\in\{9,11,13\}$ our approach successfully finds inverter-free implementations using between 7\% and 42\% fewer Majority-$3$ gates.}

\section{Introduction}

Given a technology-independent description of an application, the aim of logic synthesis is to find a hardware-specific implementation with specific properties. For example, the target functionality could be $32$-bit integer addition and the target hardware might support the gate library $\{$AND, OR, INV$\}$, which consists of two-input AND and OR gates, and inverters (INV). In addition to being correct, the implementation should have additional favorable properties that depend on the concrete use case and target hardware. A low depth and/or a low gate count are two examples of such properties.

In general, finding optimal implementations is computationally challenging already for a small number of inputs. For example, it is unknown what the optimal number of Majority-3 gates is for constructing a majority network that computes the majority of 9 inputs~\cite{testa2018mapping,soeken2018pairs}.

However, large logic networks may still be optimized heuristically. One approach is to enumerate small subnetworks that may then be optimized individually using less scalable (exact or heuristic) methods~\cite{soeken2017exact, riener2019onthefly,boyar2010new}. 
Another approach leverages local rewrites combined with, for example, simulated annealing: Starting with a correct but suboptimal logic network, the method applies various local rewrites aiming to optimize the implementation, e.g., with respect to chip area and delay~\cite{10.1145/1973009.1973095}.

The resulting networks are naturally suboptimal, and new methods that find better implementations are of great interest, as this may lead to, for example, better chip performance and lower energy consumption.

\paragraph{Contribution}

We propose a heuristic method that finds an implementation of a given function (specified, e.g., via its truth table) in terms of a given target gate library. Our method, which is based on parallel tempering Monte Carlo \cite{swendsen:86,geyer:91,marinari:92,hukushima:96}, is initialized with a random logic network that is then modified by parallel tempering updates in order to arrive at a correct implementation, i.e., one whose output agrees with that of the target function for all Boolean input vectors.

This simple approach performs surprisingly well for the benchmarks that we consider in this work, and we plan to further investigate its performance on a broader set of benchmarks in follow-up work.

\section{Logic synthesis using parallel tempering}

Given a Boolean function $f : \{0,1\}^n \rightarrow \{0,1\}^m$, we aim to find a logic network over a given gate set (for example, $\{\texttt{MAJ},\texttt{INV}\}$) that implements $f(x)$.

We propose a method based on parallel tempering Monte Carlo, which we introduce in the next subsection, before describing how we leverage parallel tempering for function synthesis in the subsequent subsection.

\subsection{Parallel tempering Monte Carlo}

Parallel tempering Monte Carlo \cite{swendsen:86,geyer:91,marinari:92,hukushima:96}
has proven to be a versatile ``workhorse'' in many fields \cite{earl:05}. The algorithm overcomes barriers in the cost function landscape by simulating several copies of the system at different temperatures. When the system swaps to a hotter temperature, variables are randomized, whereas they freeze at lower temperatures. By carefully selecting the lowest temperature to be low enough that the optimal solution manifold is reached and by keeping track of the lowest cost function values found, one can leverage parallel tempering as an effective optimizer \cite{mandra:16b}.

In more detail, $M$ non-interacting replicas of the system are simulated in parallel at different temperatures $\{T_1, T_2, \ldots, T_M\}$. After a fixed number of Monte Carlo sweeps two copies at neighboring temperatures $T_i$ and $T_{i+1}$ are exchanged with a Monte Carlo like move \cite{hukushima:96}. A given copy of the system will then perform a random walk in the temperature space. 

There are several recipes on how to ideally select the position of the temperatures for parallel tempering Monte Carlo to perform well. A measure for the efficiency of a system copy to traverse the temperature space is the probability that a swap is accepted.  A good rule of thumb is to ensure that the acceptance probabilities are approximately independent of temperature and above 20\% to ensure an effective random walk. This requirement of high enough swap rates leads to temperatures being closer together at low temperatures, which is why a geometric progression is often a good starting point \cite{PhysRevE.100.043311}.

\subsection{Description of the synthesis method}

We start by defining the energy associated with a given logic network with respect to the target function $f(x)$. Subsequently, we present the different modifications to the logic network that our method proposes in each step of the algorithm, which completes the description of our parallel tempering based synthesis method.

\subsubsection*{Energy (cost) function}
As a first step, we assign an energy $E(N)$ to a given logic network $N$ such that $E(N)=0$ if and only if the logic network $N$ implements $f(x)$ exactly, i.e., if it holds that
\[
    \forall x\in\{0,1\}^n: f(x)=N(x),
\]
where $N(x)$ denotes the output of the logic network for the input $x$.
In our experiments, we have $m=1$, i.e., we consider single-output Boolean functions, and we use
\[
    E(N) = \sum_{x\in\{0,1\}^n} f(x)\oplus N(x),
\]
where $\oplus$ denotes the logical exclusive OR of two Boolean inputs. We note that different $E(x)$ could be chosen as well, for example (and more generally for $m\geq 1$ outputs),
\[
    \tilde E(N) = \sum_{x\in\{0,1\}^n} w_x\|f(x)\oplus N(x)\|,
\]
where $w_x\in\mathds R$ are input-specific weights and $\oplus$ acts component-wise. By using appropriate weights $w_x$, \textit{don't cares} can be handled straightforwardly as well.

We want our solver to not only find a correct network $N$ implementing $f(x)$ but also to minimize the number of network nodes. Therefore, we bias the energy of logic networks depending on the number of nodes.

We start the search for a network with a maximum of $p$ nodes. If we find a network $N$ which implements $f(x)$ exactly (i.e., $E(N)=0$), we eliminate disconnected and/or trivial nodes and subnetworks to arrive at a network with $q\leq p$ nodes. We then assign the energy $E'(N) = q-p \leq 0$ to the resulting network. See table~\ref{tab:max_nodes} for the used values of $p$.

\subsubsection*{Updates}
At a given inverse temperature $\beta = 1/T$, our method proposes several updates to the logic network $N\mapsto \tilde N$, which it accepts with probability
\[
    p_A = \min(1, e^{\beta (E(N)-E(\tilde N))}).
\]
The proposed updates to $N$ may
\begin{enumerate}[1.]
    \item Assign a different input to one of the gates in $N$ \label{item:updateone}
    \item Exchange two inputs of two different gates in $N$ (if the resulting network is still valid)
    \item Randomly reassign all inputs of one of the gates in $N$
\end{enumerate}
For the results reported later, we only used update~\ref{item:updateone}. Additionally, we restrict this update to disallow trivial majority-3 gates to increase the performance of our solver. For a majority-3 gate with inputs $x_1,x_2,x_3$, denoted by $\langle x_1, x_2, x_3\rangle$, the following identities hold: $\langle x_1, x_2, x_2 \rangle \equiv x_2$ and $\langle x_1, \bar x_2, x_2 \rangle \equiv x_1$, where $\bar x_2$ denotes that the input $x_2$ is inverted. To keep the solver from exploring such trivial majority gates in a network, we require that the update assigns an input index $i$ to the majority-3 gate that is different from the other two.

\subsection{Implementation details}
Our solver implementation stores a network $N$ as a topologically sorted list of majority-3 gates. For an update, the solver chooses an input of a majority-3 gate and randomly assigns a new input that is (1) different from the other two inputs and (2) is either the output of a previous gate, a primary input, or the constant input 0 or 1.

In order to speed up the evaluation of the energy, we store the intermediate outputs of each node for all possible inputs. When a network node is updated (i.e., one of its inputs is reassigned), the solver only needs to update the intermediate outputs of the subnetwork that is affected by the change.

For each replica at temperature $T_i$ our method attempts to update each input of a gate five times before proposing to swap states at neighboring temperatures. We use an odd-even sort style swapping mechanism among the different replicas.

For our experiments, we choose the temperatures for the different replicas manually by combining two inverse temperature schedules. We determine $T_1$, $T_2$, $T_k$, $T_l$, such that the estimated acceptance rate of all encountered energy-increasing  updates is roughly 99\%, 60\%, 1\%, 0.0001\% respectively. We then insert additional replicas spaced linearly in terms of their inverse temperature $\beta=1/T$ between $T_2$ and $T_k$ and between $T_k$ and $T_l$ to achieve a swap rate above 20\%. For the results reported below, we use between 41 and 61 replicas. More advanced temperature selecting schemes or dynamically adapting temperatures during runtime can further improve the performance. We focus on this simple temperature selection scheme to showcase that our method performs well even without fine-tuning simulation parameters.

\section{Results for Majority-n}

We apply our parallel tempering solver to the task of finding implementations of Majority-$n$ (for odd $n$), which outputs 1 if and only if at least $\frac{n+1}2$ of its $n$ Boolean inputs are 1, using the elementary gate sets $\{$\texttt{MAJ}$\}$ and $\{\texttt{MAJ}, \texttt{INV}\}$, corresponding to majority networks and majority-inverter graphs (MIGs)~\cite{amaru2015majority}, respectively. \texttt{MAJ} is the three-input majority function that outputs 1 if and only if at least two of its three Boolean inputs are 1 and \texttt{INV} inverts a single Boolean input.

\subsection{Solutions without inverters}
We first consider the gate set $\{\texttt{MAJ}\}$, i.e., the inverter-free case. We compare our results to those of previous work~\cite{testa2018mapping} which, to the best of our knowledge, are the best results known to date, except for $n=9$, for which a construction is known that uses 14 \texttt{MAJ} gates~\cite{msoeken2023personal}. For small values of $n\in\{5,7\}$, exhaustive enumeration is feasible and the results are thus known to be optimal~\cite{soeken2017exact}. For $n>7$, the best known implementations were found using binary decision diagrams (BDD)~\cite{testa2018mapping}.

Our method successfully reproduces optimal majority networks for $n\in\{5,7\}$ and outperforms the BDD-based solution from~\cite{testa2018mapping} for larger values of $n$, see Table~\ref{tab:majmaj}.

\begin{table}[t]
    \centering
    \begin{tabular}{ccccccc}
        \toprule
        $n$ & 3 & 5 & 7 & 9 & 11 & 13 \\\midrule
        State of the art~\cite{testa2018mapping,msoeken2023personal} & 1 & 4 & 7 & 14 & 35 & 48 \\
        PT (this work) & 1 & 4 & 7 & \textbf{13} & \textbf{20} & \textbf{28} \\

        \bottomrule
    \end{tabular}
    \caption{Number of \texttt{MAJ} gates in the best implementation of \texttt{MAJ-n} found using our parallel tempering (PT) method compared to the state of the art from previous work~\cite{testa2018mapping}. Numbers in bold are strictly better than the previous state of the art.}
    \label{tab:majmaj}
\end{table}

\subsection{Solutions with inverters}
For our second benchmark, we allow inverters in addition to \texttt{MAJ} gates. Again, our method successfully reproduces the previously best results for $n\in\{5,7,9\}$ and it produces better results for larger $n$, see Table~\ref{tab:migmaj}. As expected, the number of \texttt{MAJ} gates required to implement Majority-$n$ does not increase compared to inverter-free solutions in Table~\ref{tab:majmaj} when allowing for inverters.

\subsection{Leafy solutions}
Our third benchmarking task is to find \textit{leafy} MAJ networks and MIGs for Majority-$n$, i.e., logic networks where at least one input of each \texttt{MAJ} gate must be a primary input~\cite{testa2018mapping}. Our method successfully finds such logic networks for $n=9$ using 13 \texttt{MAJ} gates with inverters, and using 14 \texttt{MAJ} gates without inverters.

We present the two solutions in the appendix, see Sec.~\ref{sec:networks}.

\begin{table}[t]
    \centering
    \begin{tabular}{ccccccc}
        \toprule
        $n$ & 3 & 5 & 7 & 9 & 11 & 13 \\\midrule
        State of the art~\cite{msoeken2023personal} & 1 & 4 & 7 & 12 & n/a & n/a \\
        PT (this work) & 1 & 4 & 7 & 12 & 16 & 24 \\
        \bottomrule
    \end{tabular}
    \caption{Number of \texttt{MAJ} gates in the best $\{\texttt{MAJ},\texttt{INV}\}$-implementation of \texttt{MAJ-n} found using our parallel tempering (PT) method compared to the state of the art~\cite{msoeken2023personal}.}
    \label{tab:migmaj}
\end{table}

\section{Conclusion and outlook}

We have described a Markov chain Monte Carlo method for logic synthesis that successfully finds better implementations of Majority-$n$ than previous synthesis methods.

These promising results and the general applicability of our method motivate us to investigate Markov chain Monte Carlo methods for logic synthesis more broadly. In future work, we plan to investigate the performance of such methods for other elementary gate libraries, e.g., $\{\texttt{XOR},\texttt{AND},1\}$, and for vector-valued Boolean functions.

Additionally, while our method would suffer from impractically long running times for more than about 15 inputs, it may still be used to optimize larger logic networks by resynthesizing subnetworks. It would be interesting to see whether Markov chain Monte Carlo methods yield improvements that, ultimately, translate to practical benefits such as reduced delay and/or lower chip area for specific functional units that are most relevant in practice. In such follow-up work, the effect of using don't-cares---which are also straightforward to handle in our framework---may also be investigated.

Finally, the post-processing step of our method only applies a few straightforward optimizations to the logic networks such as removing disconnected components. Optimizing these networks using other methods and tools may thus further improve the solution quality.

\section*{Acknowledgments}
We thank Gili Rosenberg for helpful comments and feedback, and Mathias Soeken for providing updated numbers for the Majority-$n$ decomposition problem.

\bibliographystyle{unsrt}
\bibliography{refs,helrefs}

\begin{thebibliography}{10}

\bibitem{testa2018mapping}
Eleonora Testa, Mathias Soeken, Luca~G Amaru, Winston Haaswijk, and Giovanni
  De~Micheli.
\newblock Mapping monotone boolean functions into majority.
\newblock {\em IEEE Transactions on Computers}, 68(5):791--797, 2018.

\bibitem{soeken2018pairs}
Mathias Soeken, Eleonora Testa, Alan Mishchenko, and Giovanni De~Micheli.
\newblock Pairs of majority-decomposing functions.
\newblock {\em Information Processing Letters}, 139:35--38, 2018.

\bibitem{soeken2017exact}
Mathias Soeken, Luca~Gaetano Amaru, Pierre-Emmanuel Gaillardon, and Giovanni
  De~Micheli.
\newblock Exact synthesis of majority-inverter graphs and its applications.
\newblock {\em IEEE Transactions on Computer-Aided Design of Integrated
  Circuits and Systems}, 36(11):1842--1855, 2017.

\bibitem{riener2019onthefly}
Heinz Riener, Winston Haaswijk, Alan Mishchenko, Giovanni De~Micheli, and
  Mathias Soeken.
\newblock On-the-fly and {DAG}-aware: Rewriting boolean networks with exact
  synthesis.
\newblock In {\em 2019 Design, Automation \& Test in Europe Conference \&
  Exhibition (DATE)}, pages 1649--1654, 2019.

\bibitem{boyar2010new}
Joan Boyar and Ren{\'e} Peralta.
\newblock A new combinational logic minimization technique with applications to
  cryptology.
\newblock In {\em Experimental Algorithms: 9th International Symposium, SEA
  2010, Ischia Island, Naples, Italy, May 20-22, 2010. Proceedings 9}, pages
  178--189. Springer, 2010.

\bibitem{10.1145/1973009.1973095}
Petra F\"{a}rm, Elena Dubrova, and Andreas Kuehlmann.
\newblock Integrated logic synthesis using simulated annealing.
\newblock In {\em Proceedings of the 21st Edition of the Great Lakes Symposium
  on Great Lakes Symposium on VLSI}, GLSVLSI '11, page 407–410, New York, NY,
  USA, 2011. Association for Computing Machinery.

\bibitem{swendsen:86}
R.~H. Swendsen and J.-S. Wang.
\newblock Replica {M}onte {C}arlo simulation of spin-glasses.
\newblock {\em Phys. Rev. Lett.}, 57:2607, 1986.

\bibitem{geyer:91}
C.~Geyer.
\newblock {Monte Carlo Maximum Likelihood for Dependent Data}.
\newblock In E.~M. Keramidas, editor, {\em 23rd Symposium on the Interface},
  page 156, Fairfax Station, VA, 1991. Interface Foundation.

\bibitem{marinari:92}
E.~Marinari and G.~Parisi.
\newblock Simulated tempering: A new {M}onte {C}arlo scheme.
\newblock {\em Europhys. Lett.}, 19:451, 1992.

\bibitem{hukushima:96}
K.~Hukushima and K.~Nemoto.
\newblock Exchange {M}onte {C}arlo method and application to spin glass
  simulations.
\newblock {\em J. Phys. Soc. Jpn.}, 65:1604, 1996.

\bibitem{earl:05}
D.~J. Earl and M.~W. Deem.
\newblock {{Parallel Tempering: Theory, Applications, and New Perspectives}}.
\newblock {\em Phys. Chem. Chem. Phys.}, 7:3910, 2005.

\bibitem{mandra:16b}
S.~{Mandr{\`a}}, Z.~{Zhu}, W.~{Wang}, A.~{Perdomo-Ortiz}, and H.~G.
  {Katzgraber}.
\newblock {Strengths and weaknesses of weak-strong cluster problems: A detailed
  overview of state-of-the-art classical heuristics versus quantum approaches}.
\newblock {\em Phys. Rev. A}, 94:022337, 2016.

\bibitem{PhysRevE.100.043311}
Ignacio Rozada, Maliheh Aramon, Jonathan Machta, and Helmut~G. Katzgraber.
\newblock Effects of setting temperatures in the parallel tempering monte carlo
  algorithm.
\newblock {\em Phys. Rev. E}, 100:043311, Oct 2019.

\bibitem{amaru2015majority}
Luca Amaru, Pierre-Emmanuel Gaillardon, and Giovanni De~Micheli.
\newblock Majority-inverter graph: A new paradigm for logic optimization.
\newblock {\em IEEE Transactions on Computer-Aided Design of Integrated
  Circuits and Systems}, 35(5):806--819, 2015.

\bibitem{msoeken2023personal}
Mathias Soeken.
\newblock Personal communication, 2023.

\end{thebibliography}

\appendix
\allowdisplaybreaks
\section{Logic formulae and simulation parameters}\label{sec:networks}
\begin{figure}[h]
    \includegraphics[width=\linewidth]{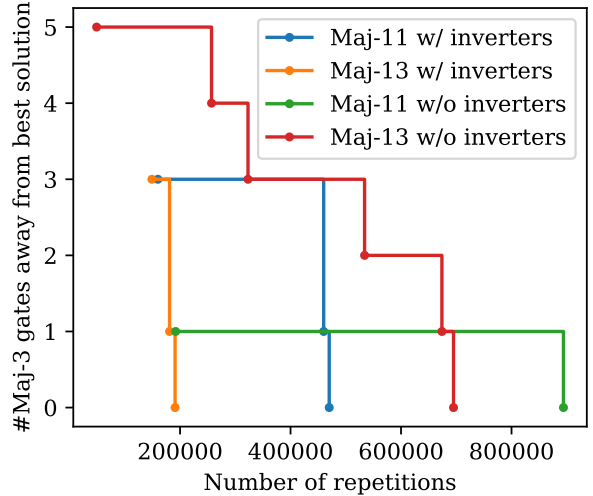}
    \caption{Best feasible solution found as a function of the number of repetitions (a repetition consists of local updates and one swap between temperatures of parallel tempering Monte Carlo). The best solutions are recorded in table~\ref{tab:majmaj} and table~\ref{tab:migmaj}.}
    \label{fig:timing_fig}
\end{figure}

Here we report the solutions found for sizes $n=9$ and larger from Table~\ref{tab:majmaj} and Table~\ref{tab:migmaj}. The time to find the solutions are in table~\ref{tab:timing}. The solver finds feasible solutions with more \texttt{MAJ} gates earlier, see figure~\ref{fig:timing_fig}, which shows that a first valid logic network was found after less than 200000 repetitions (of local updates + swaps). The runtime of a single repetition increases as the number of inputs $n$ is increased, asymptotically as $\mathcal O(2^n)$.

\begin{table}[h]
    \centering
    \begin{tabular}{cccc}
        \toprule
        $n$ & 9 & 11 & 13 \\\midrule
        without inverters & 21s & 7h34min & 41h50min  \\
        with inverters & 56s & 4h10min & 14h20min  \\
        \bottomrule
    \end{tabular}
    \caption{Runtime of our code to find the solutions in table~\ref{tab:majmaj} and table~\ref{tab:migmaj}. Note that these times depend on the random engine and seed. We run the simulation for each problem size only once and report the time it took to find the solution. We did not attempt to optimize the simulation parameters for speed. We ran on an AWS EC2 instance of type \emph{c6i.8xlarge} which has 32 vCPUs and runs on an Intel Xeon Platinum 8375C CPU.}
    \label{tab:timing}
\end{table}

\begin{table}[h]
    \centering
    \begin{tabular}{cccc}
        \toprule
        $n$ & 9 & 11 & 13 \\\midrule
        $p$ (with inverters) & 16 & 25 & 35  \\
        $p$ (without inverters) & 17 & 31 & 44  \\
        \bottomrule
    \end{tabular}
    \caption{We used a maximum number of nodes $p$ for the different instances reported in table~\ref{tab:majmaj} and table~\ref{tab:migmaj}.}
    \label{tab:max_nodes}
\end{table}

\subsection{Solutions with inverters}
Majority-9 with inverters:
\begin{align*}
    x_{0}, \dots, x_{8} &= \textrm{primary inputs} \\x_{9} &= \langle \bar x_{5} \bar x_{1} \bar x_{2} \rangle \\  x_{10} &= \langle x_{7} x_{4} x_{0} \rangle \\  x_{11} &= \langle \bar x_{6} x_{3} x_{8} \rangle \\  x_{12} &= \langle x_{5} \bar x_{1} \bar x_{2} \rangle \\  x_{13} &= \langle x_{11} x_{6} x_{10} \rangle \\  x_{14} &= \langle x_{10} \bar x_{0} \bar x_{7} \rangle \\  x_{15} &= \langle \bar x_{3} \bar x_{6} \bar x_{8} \rangle \\  x_{16} &= \langle x_{6} x_{11} \bar x_{10} \rangle \\  x_{17} &= \langle \bar x_{12} x_{5} x_{13} \rangle \\  x_{18} &= \langle x_{15} \bar x_{10} x_{9} \rangle \\  x_{19} &= \langle x_{16} \bar x_{14} x_{4} \rangle \\  x_{20} &= \langle \bar x_{18} x_{17} x_{19} \rangle
\end{align*}

Majority-11 with inverters:
\begin{align*}
    x_{0}, \dots, x_{10} &= \textrm{primary inputs} \\x_{11} &= \langle x_{2} x_{7} \bar x_{5} \rangle \\  x_{12} &= \langle x_{6} x_{1} \bar x_{0} \rangle \\  x_{13} &= \langle \bar x_{5} \bar x_{7} \bar x_{2} \rangle \\  x_{14} &= \langle x_{13} x_{11} x_{5} \rangle \\  x_{15} &= \langle \bar x_{6} \bar x_{1} \bar x_{0} \rangle \\  x_{16} &= \langle x_{10} x_{8} x_{4} \rangle \\  x_{17} &= \langle \bar x_{14} \bar x_{3} x_{9} \rangle \\  x_{18} &= \langle \bar x_{13} x_{16} \bar x_{15} \rangle \\  x_{19} &= \langle x_{10} \bar x_{8} x_{4} \rangle \\  x_{20} &= \langle x_{14} x_{9} x_{3} \rangle \\  x_{21} &= \langle x_{17} \bar x_{9} \bar x_{18} \rangle \\  x_{22} &= \langle \bar x_{16} x_{13} \bar x_{20} \rangle \\  x_{23} &= \langle \bar x_{12} \bar x_{0} x_{22} \rangle \\  x_{24} &= \langle x_{13} \bar x_{20} x_{15} \rangle \\  x_{25} &= \langle x_{8} x_{19} \bar x_{24} \rangle \\  x_{26} &= \langle x_{25} \bar x_{23} \bar x_{21} \rangle
\end{align*}

Majority-13 with inverters:
\begin{align*}
    x_{0}, \dots, x_{12} &= \textrm{primary inputs} \\x_{13} &= \langle \bar x_{6} \bar x_{10} \bar x_{12} \rangle \\  x_{14} &= \langle \bar x_{4} \bar x_{3} x_{11} \rangle \\  x_{15} &= \langle x_{0} x_{8} x_{9} \rangle \\  x_{16} &= \langle \bar x_{4} \bar x_{3} \bar x_{14} \rangle \\  x_{17} &= \langle \bar x_{2} \bar x_{5} \bar x_{1} \rangle \\  x_{18} &= \langle \bar x_{6} x_{10} \bar x_{12} \rangle \\  x_{19} &= \langle x_{15} \bar x_{9} \bar x_{0} \rangle \\  x_{20} &= \langle \bar x_{11} x_{14} x_{13} \rangle \\  x_{21} &= \langle \bar x_{2} \bar x_{17} \bar x_{1} \rangle \\  x_{22} &= \langle \bar x_{11} x_{14} \bar x_{13} \rangle \\  x_{23} &= \langle x_{16} x_{18} \bar x_{10} \rangle \\  x_{24} &= \langle \bar x_{21} \bar x_{15} \bar x_{5} \rangle \\  x_{25} &= \langle \bar x_{23} x_{16} \bar x_{10} \rangle \\  x_{26} &= \langle x_{24} x_{21} x_{23} \rangle \\  x_{27} &= \langle \bar x_{24} \bar x_{15} \bar x_{5} \rangle \\  x_{28} &= \langle \bar x_{13} \bar x_{27} \bar x_{22} \rangle \\  x_{29} &= \langle \bar x_{19} x_{8} x_{17} \rangle \\  x_{30} &= \langle x_{7} \bar x_{26} x_{28} \rangle \\  x_{31} &= \langle x_{28} x_{7} \bar x_{30} \rangle \\  x_{32} &= \langle \bar x_{19} \bar x_{29} x_{8} \rangle \\  x_{33} &= \langle \bar x_{18} \bar x_{25} x_{32} \rangle \\  x_{34} &= \langle x_{33} x_{31} \bar x_{26} \rangle \\  x_{35} &= \langle x_{29} \bar x_{20} \bar x_{17} \rangle \\  x_{36} &= \langle x_{34} x_{35} x_{30} \rangle
\end{align*}

\subsection{Solutions without inverters}
Majority-9 without inverters:
\begin{align*}
    x_{0}, \dots, x_{8} &= \textrm{primary inputs} \\x_{9} &= \langle x_{6} x_{3} x_{2} \rangle \\  x_{10} &= \langle x_{6} x_{4} x_{1} \rangle \\  x_{11} &= \langle x_{8} x_{5} x_{0} \rangle \\  x_{12} &= \langle x_{9} x_{10} x_{7} \rangle \\  x_{13} &= \langle x_{3} x_{7} x_{11} \rangle \\  x_{14} &= \langle x_{2} x_{3} x_{7} \rangle \\  x_{15} &= \langle x_{8} x_{0} x_{12} \rangle \\  x_{16} &= \langle x_{2} x_{13} x_{11} \rangle \\  x_{17} &= \langle x_{16} x_{4} x_{6} \rangle \\  x_{18} &= \langle x_{15} x_{12} x_{5} \rangle \\  x_{19} &= \langle x_{17} x_{1} x_{16} \rangle \\  x_{20} &= \langle x_{10} x_{14} x_{11} \rangle \\  x_{21} &= \langle x_{18} x_{19} x_{20} \rangle
\end{align*}

Majority-11 without inverters:
\begin{align*}
    x_{0}, \dots, x_{10} &= \textrm{primary inputs} \\x_{11} &= \langle x_{0} x_{8} x_{6} \rangle \\  x_{12} &= \langle x_{3} x_{5} x_{1} \rangle \\  x_{13} &= \langle x_{7} x_{9} x_{12} \rangle \\  x_{14} &= \langle x_{9} x_{11} x_{7} \rangle \\  x_{15} &= \langle x_{13} x_{10} x_{12} \rangle \\  x_{16} &= \langle x_{11} x_{14} x_{10} \rangle \\  x_{17} &= \langle x_{2} x_{16} x_{15} \rangle \\  x_{18} &= \langle x_{10} x_{7} x_{9} \rangle \\  x_{19} &= \langle x_{18} x_{12} x_{11} \rangle \\  x_{20} &= \langle x_{2} x_{11} x_{18} \rangle \\  x_{21} &= \langle x_{4} x_{15} x_{2} \rangle \\  x_{22} &= \langle x_{20} x_{4} x_{16} \rangle \\  x_{23} &= \langle x_{22} x_{1} x_{5} \rangle \\  x_{24} &= \langle x_{4} x_{19} x_{17} \rangle \\  x_{25} &= \langle x_{21} x_{18} x_{12} \rangle \\  x_{26} &= \langle x_{22} x_{3} x_{23} \rangle \\  x_{27} &= \langle x_{6} x_{8} x_{25} \rangle \\  x_{28} &= \langle x_{25} x_{0} x_{27} \rangle \\  x_{29} &= \langle x_{24} x_{2} x_{19} \rangle \\  x_{30} &= \langle x_{28} x_{29} x_{26} \rangle 
\end{align*}

Majority-13 without inverters:
\begin{align*} 
    x_{0}, \dots, x_{12} &= \textrm{primary inputs} \\x_{13} &= \langle x_{8} x_{10} x_{11} \rangle \\  x_{14} &= \langle x_{7} x_{9} x_{5} \rangle \\  x_{15} &= \langle x_{1} x_{6} x_{2} \rangle \\  x_{16} &= \langle x_{11} x_{10} x_{15} \rangle \\  x_{17} &= \langle x_{14} x_{4} x_{3} \rangle \\  x_{18} &= \langle x_{3} x_{4} x_{0} \rangle \\  x_{19} &= \langle x_{18} x_{7} x_{5} \rangle \\  x_{20} &= \langle x_{15} x_{8} x_{16} \rangle \\  x_{21} &= \langle x_{0} x_{17} x_{14} \rangle \\  x_{22} &= \langle x_{13} x_{1} x_{6} \rangle \\  x_{23} &= \langle x_{0} x_{20} x_{3} \rangle \\  x_{24} &= \langle x_{19} x_{18} x_{9} \rangle \\  x_{25} &= \langle x_{2} x_{1} x_{24} \rangle \\  x_{26} &= \langle x_{24} x_{11} x_{8} \rangle \\  x_{27} &= \langle x_{4} x_{20} x_{14} \rangle \\  x_{28} &= \langle x_{21} x_{10} x_{26} \rangle \\  x_{29} &= \langle x_{22} x_{2} x_{13} \rangle \\  x_{30} &= \langle x_{5} x_{29} x_{9} \rangle \\  x_{31} &= \langle x_{23} x_{27} x_{20} \rangle \\  x_{32} &= \langle x_{25} x_{6} x_{24} \rangle \\  x_{33} &= \langle x_{15} x_{21} x_{28} \rangle \\  x_{34} &= \langle x_{13} x_{24} x_{32} \rangle \\  x_{35} &= \langle x_{29} x_{30} x_{7} \rangle \\  x_{36} &= \langle x_{18} x_{35} x_{29} \rangle \\  x_{37} &= \langle x_{31} x_{33} x_{12} \rangle \\  x_{38} &= \langle x_{12} x_{36} x_{31} \rangle \\  x_{39} &= \langle x_{36} x_{37} x_{33} \rangle \\  x_{40} &= \langle x_{39} x_{34} x_{38} \rangle
\end{align*}

\subsection{Leafy solutions for Majority-9}
Leafy Majority-9 with inverters:
\begin{align*} 
    x_{0}, \dots, x_{8} &= \textrm{primary inputs} \\x_{9} &= \langle x_{6} x_{8} x_{3} \rangle \\  x_{10} &= \langle x_{7} x_{2} x_{1} \rangle \\  x_{11} &= \langle x_{7} \bar x_{2} \bar x_{1} \rangle \\  x_{12} &= \langle \bar x_{5} \bar x_{4} \bar x_{3} \rangle \\  x_{13} &= \langle x_{10} x_{4} x_{9} \rangle \\  x_{14} &= \langle x_{7} x_{9} \bar x_{11} \rangle \\  x_{15} &= \langle \bar x_{10} \bar x_{5} \bar x_{4} \rangle \\  x_{16} &= \langle \bar x_{13} \bar x_{14} \bar x_{5} \rangle \\  x_{17} &= \langle \bar x_{3} \bar x_{14} x_{12} \rangle \\  x_{18} &= \langle \bar x_{11} \bar x_{15} x_{7} \rangle \\  x_{19} &= \langle \bar x_{17} x_{8} \bar x_{9} \rangle \\  x_{20} &= \langle x_{6} x_{19} x_{18} \rangle \\  x_{21} &= \langle x_{0} x_{20} \bar x_{16} \rangle
\end{align*}

Leafy Majority-9 without inverters:
\begin{align*} 
    x_{0}, \dots, x_{8} &= \textrm{primary inputs} \\x_{9} &= \langle x_{1} x_{8} x_{3} \rangle \\  x_{10} &= \langle x_{9} x_{6} x_{4} \rangle \\  x_{11} &= \langle x_{0} x_{6} x_{4} \rangle \\  x_{12} &= \langle x_{0} x_{9} x_{10} \rangle \\  x_{13} &= \langle x_{0} x_{6} x_{7} \rangle \\  x_{14} &= \langle x_{11} x_{9} x_{7} \rangle \\  x_{15} &= \langle x_{5} x_{12} x_{14} \rangle \\  x_{16} &= \langle x_{11} x_{5} x_{7} \rangle \\  x_{17} &= \langle x_{14} x_{2} x_{5} \rangle \\  x_{18} &= \langle x_{16} x_{4} x_{13} \rangle \\  x_{19} &= \langle x_{17} x_{12} x_{8} \rangle \\  x_{20} &= \langle x_{19} x_{3} x_{18} \rangle \\  x_{21} &= \langle x_{20} x_{1} x_{18} \rangle \\  x_{22} &= \langle x_{15} x_{21} x_{2} \rangle
\end{align*}
\end{document}